\documentclass[aip,jcp,reprint,showpacs,twocolumn,twoside,floatfix,a4]{revtex4-1}
\usepackage{hyperref}
\usepackage{amsmath, amssymb,wrapfig}
\usepackage{color, array,soul}
\usepackage{graphicx,epsfig}
\usepackage{mathptmx} 
\usepackage{amssymb}

\renewcommand{\d}{\textrm d}
\def\H{ {\mathcal H} }

\newcommand{\bra}[1]{\langle #1|}
\newcommand{\ket}[1]{|#1\rangle}
\newcommand{\braket}[2]{\langle #1|#2\rangle}

\begin{document}

\title{A qualitative quantum rate model for hydrogen transfer in soybean lipoxygenase}

\author{S. Jevtic}
\email{sania.jevtic@imperial.ac.uk}
\affiliation{Department of Mathematics, Huxley Building, Imperial College, London SW7 2AZ, United Kingdom}
%\affiliation{{\color{red}Institut f\"{u}r Theoretische Physik, Leibniz Universit\"{a}t Hannover, Appelstra{\ss}e 2, 30167 Hannover, Germany}}
\author{J. Anders}
\email{janet@qipc.org}
\affiliation{CEMPS, Physics and Astronomy, University of Exeter, Exeter EX4 4QL, United Kingdom}

%\date{\today}

\begin{abstract}

The hydrogen transfer reaction catalysed by soybean lipoxygenase (SLO) has been the focus of intense study following observations of a high kinetic isotope effect (KIE). Today high KIEs are generally thought to indicate departure from classical rate theory and are seen as a strong signature of tunnelling of the transferring particle, hydrogen or one of its isotopes, through the reaction energy barrier.  In this paper we build a qualitative quantum rate model with few free parameters that describes the dynamics of the transferring particle when it is exposed to energetic potentials exerted by the donor and the acceptor. The enzyme's impact on the dynamics is modelled by an additional energetic term, an oscillatory contribution known as ``gating''. By varying two key parameters, the gating frequency and the mean donor-acceptor separation, the model is able to reproduce well the KIE data for SLO wild-type and a variety of SLO mutants over the experimentally accessible temperature range. While SLO-specific constants have been considered here, it is possible to adapt these for other enzymes. 

\end{abstract}

\maketitle

%%%%%%%%%%%%%%%%%%%%%%%%%
\section{\label{sec_Intro}Introduction}

Enzymes play a central role in biological functions and are indispensable in many industrial processes \cite{SBT02}. As such, there is a pressing need to develop theoretical models that fully specify the method by which enzymes catalyse reactions.  An enzyme creates an alternative path for a reaction to occur and can greatly speed up the reaction compared to the uncatalysed case; speed-up factors of up to $10^{26}$ have been observed \cite{ELW12}. As well as being able to simulate the data of known enzymes, it is crucial to find a model that can predict the action of a potential catalyst. This will enable the engineering of new enzymes for reactions that are currently too slow \cite{HB03}.

\begin{figure}[t]
\hspace{1.5cm}\includegraphics[trim=2cm 0cm 0cm 0cm, clip,width=3.2in]{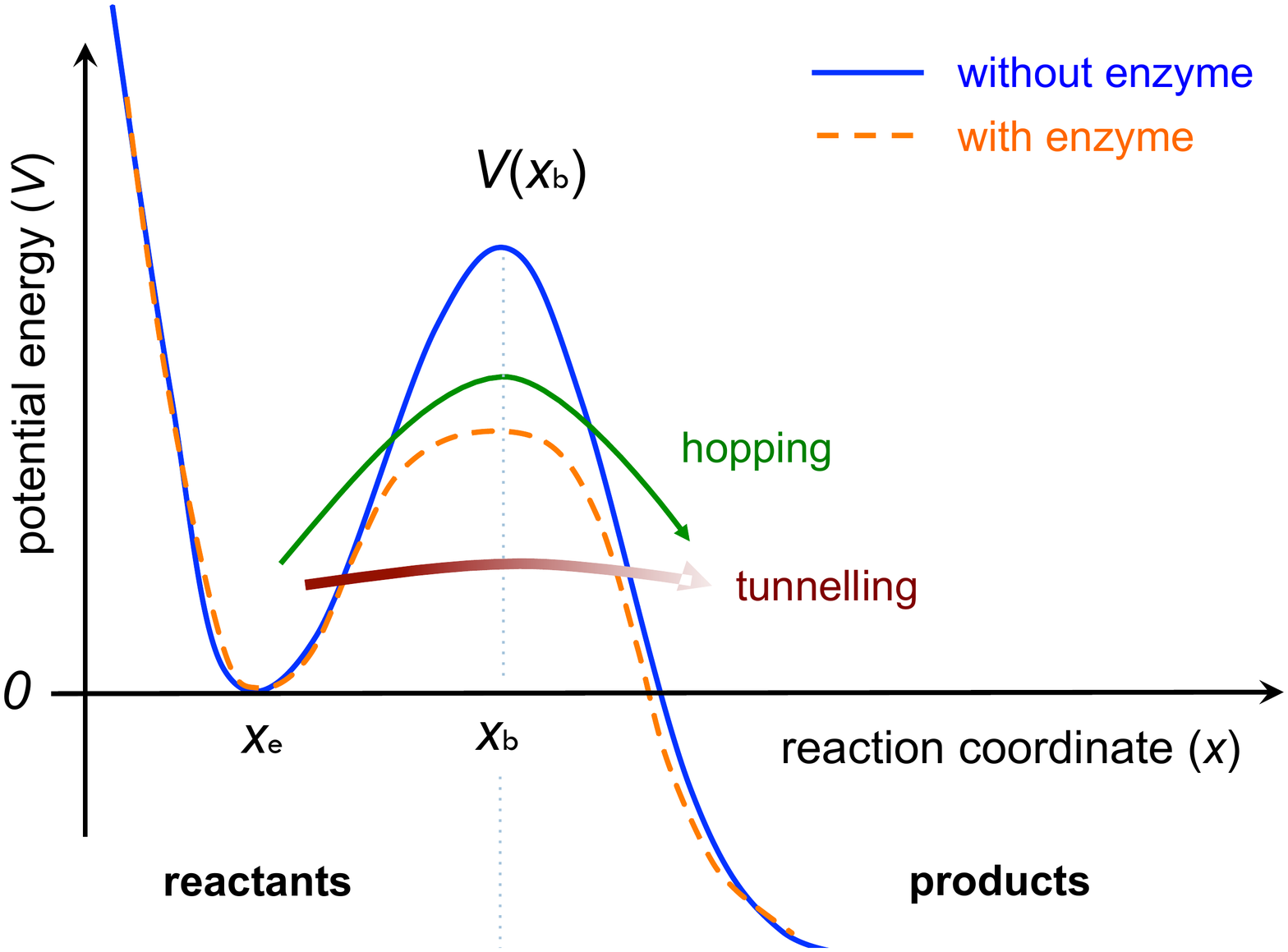}
\caption{Sketch of the potential energy profile $V(x)$ experienced by a particle transferring from a donor (reactants) to an acceptor (products). The particle, initially bonded to the donor, sits in a potential minimum at position $x_e$. The reaction proceeds along a reaction coordinate $x$ and at position $x_b$ a barrier of height $V(x_b)$ must be overcome if the particle is to break free of the donor and form a new bond with the acceptor. The particle may be thermally excited and hop over the barrier if the transfer is viewed classically, or it may tunnel through the barrier if it is governed by quantum dynamics. Enzymes are believed to catalyse the reaction by lowering the barrier height. }
\label{E_barrier}
\end{figure}

The standard method for modelling enzyme reaction rates is based on transition state theory (TST) \cite{ES39}. In TST, the reactants begin in a local minimum of a potential $V(x)$, proceed along a reaction coordinate $x$, and at $x=x_b$ they encounter an energetic barrier of height $V(x_b)$. Thermal excitations from the environment enable the formation of the transition state at the top of the barrier, and crossing the barrier leads to the products being created, see Fig.~\ref{E_barrier}.  In this picture, the catalyst lowers the energy barrier increasing the likelihood for the transition state to be formed and the transferring particle to hop over the barrier. An alternative transfer mechanism is also possible: the transferring particle may tunnel \cite{ETunnel} through the barrier instead of hopping over it. This has been discussed in a number of enzymatic systems that catalyse hydrogen transfer and have high kinetic isotope effects (KIEs), such as soybean lipoxygenase (SLO)~\cite{SHS_04,SHS_07,SHH_S_2005,ESS_H_2010, SH_S_2015,SLO_DM, SH_S_2016,KRK02, Warshel, Cha_Klinman_1989,debate,debate2,PB04,Scrutton_book,BGM10, Scrutton_book,Layfield_SHS, SS_2008}. 

In this manuscript, we present a rate model that aims to capture qualitatively the mechanism of hydrogen transfer in enzyme-catalysed reactions. The purpose of the model is to be able to predict the temperature dependence of the KIEs for various enzyme mutants, which are parameterised by a few key parameters. To achieve this, we build a rate model that treats the dynamics of the transferring particle quantum mechanically and also allow the enzyme to sample a range of donor-acceptor configurations through a classical vibrating motion, known as \textit{gating} \cite{BB92}, which arises due to thermal excitations from the environment at temperature $T$. 

For concreteness, here we focus on hydrogen and deuterium transfer catalysed by the enzyme SLO, and its mutants. We choose two independent quantities to parametrise the rate of transfer, the average donor-acceptor separation, $R_e$, and the gating frequency, $\Omega$. Our calculated KIEs show good agreement with the experimental KIE curves for wild-type (WT) SLO and four different SLO mutants reported in Refs~[\onlinecite{KRK02}, \onlinecite{SLO_DM}] over the measured temperature range of $5-50^\circ$C. The parameter choices provide insight into the physical features that affect the reaction rates and KIEs, such as the donor-acceptor configurations of SLO mutants in comparison to WT.  The model also allows us to discuss the magnitude of the tunnelling contribution. 

The paper is organised as follows. In section \ref{sec_SLOdata}, we introduce the enzyme SLO, summarise pertinent experimental results, and briefly discuss a selection of existing rate models. We present our new qualitative quantum rate model in section \ref{sec3} and discuss the key results in section \ref{sec4}. In section \ref{sec5} we comment on whether tunnelling plays a significant role in SLO enzyme catalysis. Finally, in section \ref{sec7}, we discuss insights arising from the comparison between the proposed model and the experimental data, and suggest future directions.

%%%%%%%%%
%%%%%%%%%

%%%%%%%%%%%%%%%%%%%%%%%%%
\section{\label{sec_SLOdata}The enzyme SLO}

Soybean lipoxygenase (SLO) is studied because of its similarities to the mammalian lipoxygenases. These are key components in the production of fatty acids which are required for the functioning of cells \cite{PA96,PA97}. Abnormal lipoxygenase activity has been linked with cancer formation, hence these enzymes play an important role in human health and are of particular interest to the pharmaceutical industry \cite{YF04,RC98,MM98}. SLO catalyses the production of fatty acid hydroperoxides and the substrate is linocleic acid \cite{KRK02}. The reaction consists of a sequence of rapid steps, however, the rate-limiting step is the hydrogen transfer from a carbon atom on linocleic acid to an oxygen molecule. This is the step that is modelled in the quantum dynamical rate model developed in section \ref{sec3}. 

%%%%%%%%%
\subsection{Kinetic isotope effect (KIE)}

The first clear deviation from standard enzyme kinetics was reported in the kinetic isotope effect (KIE) of soybean lipoxygenase more than 20 years ago \cite{GWK94}. The KIE is an experimental tool for testing the mass-dependence of a reaction rate. It is the ratio between two rates: the rate of hydrogen transfer, $k_H$, and the rate of transfer of one of its isotopes, e.g. deuterium, $k_D$. (When the transferring particle is substituted by one of its isotopes this is called the primary KIE, and this is the situation we consider here. Secondary KIEs refer to rate changes that occur when isotopically substituting a non-transferring particle in the reactant.) 
The isotope substitution does not affect the electrostatic potentials, however, the mass change affects the zero point energy. At 30$^{\circ}$C this can reduce the deuterium rate by a factor of 1.4 - 3 per normal mode (e.g. squeezing or bending modes) leading to increased KIEs. Experimental SLO rates, shown in Fig.~\ref{figKRK02}, exhibit a huge KIE = $k_H / k_D = 81$ at 30$^{\circ}$C.

%%%%%%%%%
\subsection{Mutations}

\begin{figure}[t]
\includegraphics[trim=1cm 0.5cm 13cm 0.8cm, clip, width=3.0in]{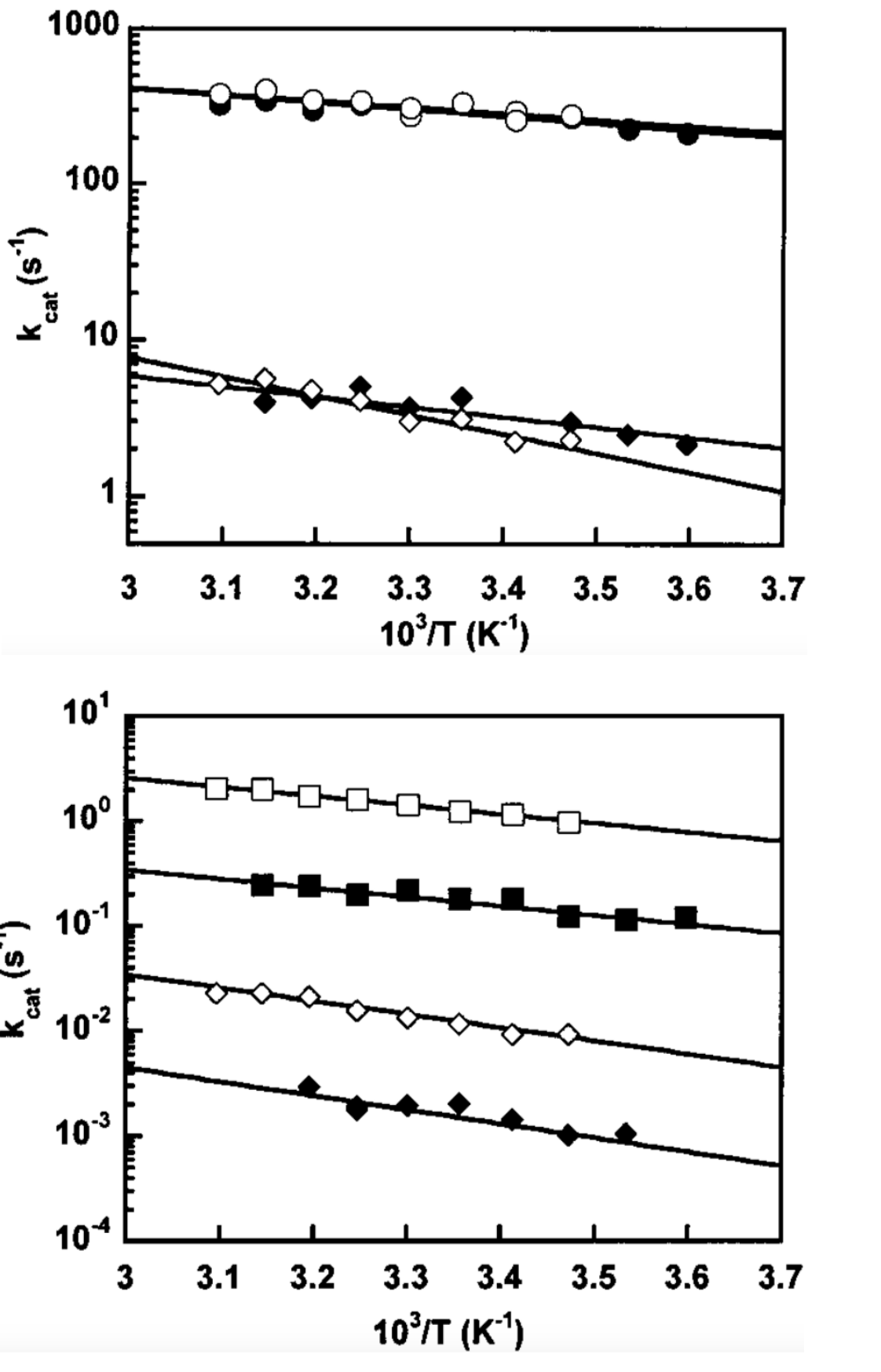}
\caption{Arrhenius plots of measured reaction rates over $10^3/T$ for SLO in the temperature $T$ range $5-50^{\circ}$C. {\bf Top:} Reaction rate data points for  SLO wild-type (black symbols) and SLO mutation M3 (Ile$^{553}\rightarrow$ Ala) (white symbols) for hydrogen transfer (circles) and deuterium transfer (diamonds). While the hydrogen transfer is not affected by mutation M3, the deuterium transfer is affected and this leads to a higher KIE for M3 than for WT. The solid lines are non-linear fits to the Arrhenius equation. 
{\bf Bottom:} Rates for SLO mutation M1 (Leu$^{546} \rightarrow$ Ala) (white symbols) and M2 (Leu$^{754} \rightarrow$ Ala) (black symbols) for hydrogen (squares) and deuterium (diamonds) transfer. The data show that mutants M1 and M2 have significantly lower rates than WT, i.e. they are much less efficient in catalysing the reaction. Reprinted with permission from J. Am. Chem. Soc. 124, 3865 (2002). Copyright 2002 American Chemical Society.}
\label{figKRK02}
\end{figure}

Aside from deuterating the transferring particle, it is possible to \textit{mutate} the enzyme by substituting large clusters of atoms (``residues") with smaller ones \cite{KRK02,MTK08,SLO_DM}. This is carefully done so that the enzyme catalyses the same reaction but the rate is altered. In [\onlinecite{KRK02}], several ``bulky'' residues (leucine (Leu) 546 and 754, and isoleucine (Ile) 553) near the active site of SLO are replaced by the smaller amino acid alanine (Ala). Such mutations modify the active site and so hydrogen will be exposed to a different potential energy barrier. For the mutations Leu$^{546} \rightarrow$ Ala (mutation M1) or Leu$^{754} \rightarrow$ Ala (mutation M2), which are both close to the active site, both rates $k_H$ and $k_D$ significantly drop (about 3 orders of magnitude) in comparison to WT SLO, see Fig.~\ref{figKRK02}. These findings indicate that wild-type SLO is configured optimally to catalyse this reaction. The KIEs of mutants M1 and M2 are larger than WT, 109 and 112 respectively at 30$^\circ$C, and show stronger variation with temperature. In contrast, the more distant mutation Ile$^{553}\rightarrow$ Ala (mutation M3) barely changes the rate $k_H$, see Fig.~\ref{figKRK02}, but the M3 KIE is more temperature dependent than the KIEs of WT and mutations M1 and M2. These observations were confirmed once more in [\onlinecite{MTK08}]. Recently, kinetic data for the SLO \textit{double mutant} (DM) have been obtained \cite{SLO_DM}. The double mutation makes both replacements M1 and M2 at the same time in SLO. Using two independent experimental methods, hugely inflated KIEs were observed: a KIE of $537\pm 55$ at $35^\circ$C was measured using single-turnover kinetics and a KIE of $729\pm 26$ at $30^\circ$C was measured using steady-state measurements.

%%%%%%%%%
%%%%%%%%%

\subsection{Advanced models of enzyme catalysis}

It is widely accepted that enzymes with high KIEs, such as SLO WT and its mutants, require quantum corrections, such as the inclusion of thermally activated tunnelling \cite{CON77,Bell_Tunnel,Scrutton_book}. Other quantum corrections include making the Wentzel-Kramers-Brillouin (WKB) approximation that treats tunnelling semiclassically \cite{BB92}. 
However, for transfer distances of $1~\AA$ and activation energies of $10^{-20}$J in SLO, this approximation is not fully justified \footnote{The WKB approximation provides a good estimate of the tunnelling probability through a barrier when $\hbar/S \ll 1$. In this case the tunnelling probability is $\propto \exp(-2S/\hbar)$. The action $S = \int_a^b \sqrt{2m[V(x)-E]}dx$ is for a one-dimensional potential $V(x)$ that has a barrier in the range $a \leq x \leq b$, and the incident particle has energy $E$ and mass $m$. For the purposes of an order estimation, let us assume the barrier is rectangular, i.e. $V(x) = V$ for $a \leq x \leq b$ and zero otherwise. Then $S = (b-a)\sqrt{2m[V-E]}$. Typical parameters for the transfer distance $b-a \sim 1~\AA$ and for the activation energy $V-E \sim 1$ kcal/mol $\sim 1\times 10^{-20} $ J (from SLO data in [\onlinecite{KRK02}]). Hence $\hbar/S  \sim 0.1$ indicating that the WKB regime is not a very good approximation.}. Nevertheless, semiclassical rate theories have been successful in simulating a variety of non-classical enzymes \cite{BGM10}, including SLO \cite{Pollak}. 

An array of quantum rate models has been developed that account for the increased complexity of the rate-determining step in SLO \cite{KU99, KRK02, Warshel, Mincer_Schwartz, SS04,IS2008,IS2010} for the WT and mutants M1, M2 and M3.  A successful framework employing Fermi's golden rule is presented by Hammes-Schiffer and co-workers \cite{SHS_AS_2000,SHS_04,SHH_S_2005,SHS_07,SS_2008,ESS_H_2010, AS_SHS_2014,SLO_DM, SH_S_2015, SH_S_2016}. This rate model provides a good fit to the wild-type SLO KIE data\cite{KRK02} as well as predicting the KIE magnitude and temperature dependence of the mutants M1, M2, M3 (and its variants \cite{MTK08}) and the double mutant \cite{SLO_DM}. This approach incorporates hydrogen transfer into Marcus theory through ``proton-coupled electron transfer'' and combines this with \textit{gating}.  
Gating is the sampling of different configurations of the active site including close confinement where quantum tunnelling is possible \cite{BB92}. This sampling is caused by the enzyme's thermal vibrating motion which reorganises the active site and modulates the barrier.

An intensive computational study was carried out in [\onlinecite{TVLLY06}] using ensemble averaged variational TST with multidimensional tunnelling \cite{PGT06} to calculate the SLO rate and KIE. The authors reported that hydrogen tunnelling accounted for over 99\% of the transfer mechanism in the wild type setting. The KIEs they obtain ($\sim10$) are far lower than the observed value of 81. It is believed that this is due to an underestimation of the barrier height which comes from their computed hydrogen potential energy surface. Manually increasing the barrier height (and width) rapidly leads to an increased KIE. 

The above quantum rate models have been shown to match the observed KIE data. However, many of them are rather complex and require the fixing of numerous parameters. Rates calculated for different parameter choices are checked for consistency with the data, but the complexity of how the parameters affect the rates could limit the models' ability to make predictions for new experiments. We note that apart from the  quantum models mentioned above, a semiclassical model, which leads to a Langevin equation including friction, has also shown agreement with the experimental data\cite{Pollak}. While the individual rates are not specified, this model requires only a single parameter for each mutant, the friction coefficient, to obtain the corresponding KIE curves. 

Here we aim to develop a quantum rate model with limited complexity (two parameters) that qualitatively produces the observed KIEs and temperature variation for various mutants. To benchmark the proposed model we will compare its predictions with the conclusions drawn from another two-parameter model that has previously been discussed \cite{ESS_H_2010,SH_S_2016}. 

%%%%%%%%%%%%%
\section{\label{sec3} A qualitative quantum model with classical gating}

Building on previous rate models we propose here a new qualitative model for enzyme-catalysed hydrogen transfer that treats the dynamics of the transferring particle fully quantum mechanically. The model does not make semiclassical approximations, such as WKB. Instead the model calculates coherent quantum dynamics contributions to the rate. These contributions are then averaged over the active site configurations which are sampled by the enzyme's vibration (gating).

%%%%%%%%%%%%%
\subsection{Overview of the rate model} 

The model assumes that hydrogen, or one of its isotopes, is initially in thermal equilibrium in a potential $V^C$ created by the donor atom (carbon, C). When the  acceptor atom (oxygen, O) is brought close by the enzyme, the hydrogen experiences a different potential, $V^{CO}_R$, which is parametrised by the donor-acceptor separation $R$.  The hydrogen atom is in a non-stationary state with respect to the new potential $V^{CO}_R$ and this will result in quantum dynamics with the state of hydrogen evolving according to the Schr\"odinger equation. We obtain a quantum rate $\tau_R$ that quantifies the rate of the hydrogen transferring from the donor to the acceptor for each value of $R$. To obtain a prediction of the experimentally measured rate these quantum rates are then weighted with a classical gating probability $p(R)$ that determines the likelihood of the donor-acceptor distance $R$ being realised in a thermal environment \cite{BB92}. The variation of this distance over a range $R_i \leq R \leq R_f$ is realised by the enzyme vibration, i.e. ``gating''.  Averaging over the range of $R$ then gives the overall transfer rate,
\begin{eqnarray}
	k= \frac{N}{|R_f - R_i|}\int^{R_f}_{R_i} p(R)\, \tau_R \, \d R.
\label{Our_avg_k}
\end{eqnarray}
Here $N$ is a dimensionless prefactor that accounts for factors that will influence the experimentally measured rate, but do not directly relate to the particle transfer assisted by the enzyme in the rate-limiting step. These factors include the probability of the reactants coming together in the active site in the first place, as well as any other relevant effects due to the environment, for instance the concentration of the solvent. Thus $N$ will be temperature dependent although we expect it to have a significantly weaker temperature dependence than the temperature dependence of the other factor in the rate expression, $\frac{1}{|R_f - R_i|}\int^{R_f}_{R_i} p(R)\, \tau_R \, \d R$. We also assume that $N$ is independent of the mass of the transferring particle, i.e. it is the same for all isotopes. This is justified because the charge of the transferring particle, which may cause long-range interactions with the environment, is constant for all isotopes.

This rate expression allows one to predict the mass and temperature dependence of ratios of rates, i.e. KIEs, where environmental effects captured in the prefactor $N$ cancel out. The thermal vibrations of the enzyme influence the reaction by realising a configuration where the donor and acceptor are at a distance $R$ with probability $p(R)$. This probability will be a function of temperature and is determined by two parameters: the mean donor-acceptor distance $R_e$ and the gating frequency $\Omega$ of donor-acceptor oscillations about their equilibrium separation realised by the enzyme. The quantum rate $\tau_R$ derives from the Schr\"{o}dinger equation of the transferring particle and is thus mass-dependent. In principle, $\tau_R$ is weakly temperature dependent, too, because of thermal occupation of the C-H bond energies before catalysis. However, as we will see this dependence is negligible at biological temperatures. 

Choosing a specific hydrogen isotope, and thus the mass, fixes $\tau_R$. The only variables left to obtain the KIEs for each SLO variant are then $R_e$ and $\Omega$, which determine the gating probability $p(R)$. The following subsections will discuss the factors $\tau_R$ and $p(R)$ and their parameter dependence in more detail. 

%%%%%%%%%%
\begin{figure}[t]
\includegraphics[trim={8cm 9.2cm 9.2cm 3.2cm}, clip,width=0.40\textwidth]{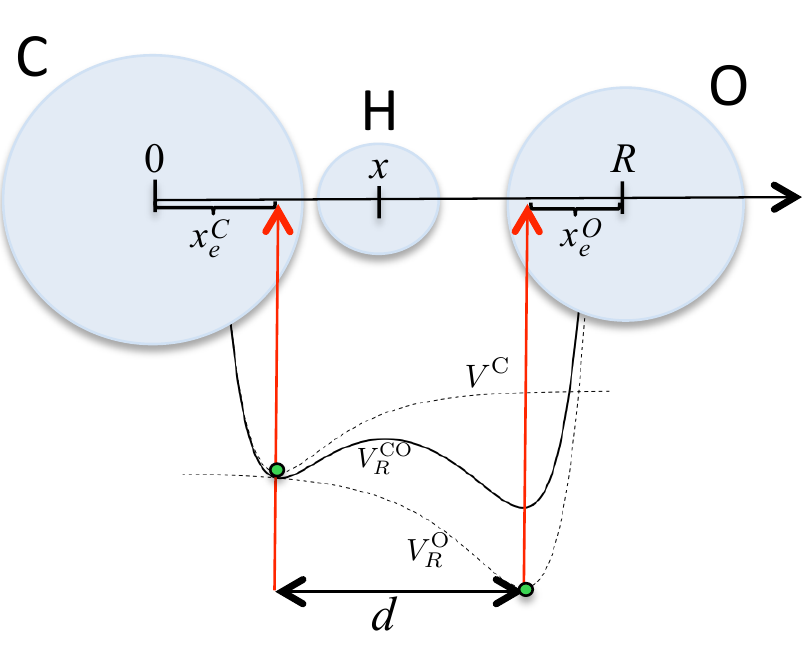} 
\caption{The configuration of the donor carbon (C), transferring hydrogen (H) and acceptor oxygen (O) atoms. $x$ denotes the separation between H and C, while $R$ denotes the separation of O from C. The equilibrium internuclear separations of C-H ($x_e^C$) and O-H ($x_e^O$) are shown. They define the positions of the minima of the C-H and O-H Morse potentials, $V^C$ and $V_R^O$ (dashed black curves). The wells of $V^C$ and $V_R^O$ are separated by a distance $d$ and $V^C$ and $V_R^O$ are summed to give the asymmetric double well potential $V_R^{CO}$ (solid black line) experienced by H.}
\label{fig_cho_config}
\end{figure}
%%%%%%%%%%

%%%%%%%%%%%%%
\subsection{The quantum mechanical transfer rate $\tau_R$} 

Prior to the transfer event, hydrogen (H) or one of its isotopes, is bonded to the donor carbon (C) atom and in thermal equilibrium with its environment at temperature $T = (k_B \beta)^{-1}$.  The initial state of hydrogen is thus the stationary, thermal state $\rho = \sum_{n=0}^{\infty}  \ket{E^C_n}\bra{E^C_n} \, e^{- \beta E^C_n}/Z^C_{\beta}$ of the C-H interaction potential $V^C$, where $E^C_n$ are the eigenenergies and $\ket{E^C_n}$ the energy eigenstates of $V^C$, and $Z^C_{\beta}$ is the partition function. However, in the biological temperature range $T \in [5^\circ\mbox{C}, 50^\circ\mbox{C}]$ the parameters that determine $V^C$, discussed in subsection~\ref{secParams}, result in excited energy levels of $V^C$ that are too high to be significantly populated. Thus the probability of hydrogen occupying the ground state $\ket{E_0^{\mathrm{C}}}$ of the potential $V^\mathrm{C}$ is over $99\%$.  Therefore the rate $\tau_R$ will be determined solely by the ground state evolution and so is temperature-independent. 

When the enzyme brings the donor and acceptor atoms into close confinement in its active site, with donor-acceptor distance $R$, the hydrogen becomes exposed to an asymmetric double well potential $V_R^{CO}$ due to its interaction with the nearby acceptor oxygen (O) atom. Since the potential is suddenly changed, the initial state (ground state $\ket{E^C_0}$ of  $V^C$) is no longer a stationary state for the new potential $V_R^{CO}$ and so the probability of finding hydrogen near the acceptor changes over time.

Assuming $V_R^{CO}$ is constant during the small transfer time window $t_{\max}$, then the state of the transferring hydrogen atom at intermediate times $t \in [0, t_{\max}]$ is
\begin{equation}
	\ket{\psi_R(t)} =\exp \left( -i  \H_R^{CO}t / \hbar \right) \ket{E^C_0},
\end{equation}
where $\H_R^{CO} =\frac{p^{2}}{2m}+V_R^{CO}$ is the Hamiltonian that generates the evolution from the initial ground state $\ket{E^C_0}$ and  $m$ is the mass of the transferring particle. The potential $V_R^{CO}$ will be a double well potential for larger values of $R$, see Fig.~\ref{fig_cho_config}, with further details for $V_R^{CO}$ described below. The probability of hydrogen transfer is the probability of observing it on the acceptor site at time $t$, $\varphi_R(t)= \int_{x_b}^{\infty} |\braket{x}{\psi_R(t)}|^2 \, \d x$, i.e. the hydrogen is anywhere to the right of the barrier peak position $x_b$, see Fig.~\ref{E_barrier}. If there is no barrier, which can occur when C and O are very close, then $V_R^{CO}$ has a single well and $x_b$ is defined as the position of the minimum of $V_R^{CO}$. Physically this means that, at large donor-acceptor distances, the transferring particle is strongly localised either at the donor or the acceptor, while at smaller distances it is shared between the two.

To obtain a rate constant we average the time-derivative of $\varphi(t)$ over a time window $t_{\max}$. We choose this time window as the smallest timescale on which thermal relaxation will  affect the Schr\"odinger evolution of the system. This damping time is given by $t_{\max} = \frac{\hbar}{\Delta E}$ where $\Delta E$ is the energy gap\cite{BP_OpenQS} that the transferring particle sees at the donor, i.e. the energy gap between the ground and first excited states of the C-H bond (or C-D bond for deuterium).
The Schr\"odinger evolution of the hydrogen atom then gives rise to the rate
\begin{eqnarray}
\label{eqn_tau}
	\tau_R &=&\frac{1}{t_{\max}}\int_{0}^{t_{\max}} {\d {\varphi}_R(t) \over \d t} \, \d t.
\end{eqnarray}

To calculate the rate $\tau_R$ we  now detail the potential $V_R^{CO}$. It is composed of the two Morse potentials seen by the transferring particle due to the presence of the donor  and acceptor. The Morse potential has the form $V^Y(x) = D^Y \left(1-e^{- g \, a^Y(x-x^Y_e)}\right)^2$ for each Y-H bond where Y is either C (donor) or O (acceptor). Here $D^Y$ is the well depth, $x^Y_e$ is the equilibrium separation and $a^Y = \omega^Y\sqrt{\mu^Y/2D^Y}$ is the well ``curvature''.  $\mu^Y$ is the reduced mass between hydrogen and Y, and $\omega^Y$ is the bond frequency. These constants can be found in the literature for the case when only two particles, either C-H or O-H, are bonded. To account for the fact that the transferring particle before (after) the transfer sees the electrostatic potential not just of a single carbon (oxygen) atom, but of these atoms when part of donor (acceptor) molecule, the squeezing parameter $g$ has been introduced in $V^Y(x)$. 

Assuming that the donor, hydrogen, and acceptor atoms are collinear in a single reaction coordinate $x$, see Fig.~\ref{fig_cho_config}, the combined potential seen by hydrogen at C-O separation $R$ is $V_R^{CO}(x) = V^C(x) + V_R^O(x) - D^O$,  obtained by summing $V^C(x) = D^C \left((1-e^{-ga^C(x-x^C_e)}\right)^2$ and $V_R^O(x) = D^O \left(1-e^{-ga^O(-x+(R-x_e^O))}\right)^2$ together with an offset $-D^O$. This offset guarantees that, if the acceptor (O) were moved infinitely far away from the donor (C), the hydrogen would feel no force due to the acceptor. The C-O separation is $R = d + x_e^O + x_e^C$, where $x_e^O$ and $x_e^C$ are fixed, and $d$ can vary, see Fig.~\ref{fig_cho_config}. Note that $d$ defines the separation between the $V^C$ and $V^O_R$ well minima which can deviate from the well-separation of the two wells in the resulting $V^{CO}_R$. 

While isotopes of hydrogen will have a different mass from hydrogen, this mass has no effect on the geometry of the system and on the electrostatic forces involved. Consequently, the potentials $V^C, V_R^O, V_R^{CO}$  remain the same for all isotopes. However, the eigenenergies and eigenstates of the corresponding Hamiltonians $\H^C, \H_R^O, \H_R^{CO}$ are mass-dependent, because mass enters the Schr\"odinger equation through the kinetic term $\frac{p^2}{2m}$. This is what makes the quantum contribution $\tau_R$, and therefore the overall rate $k$, dependent on the mass of the transferring particle. Denoting the hydrogen rate by $k_H$ and the deuterium rate by $k_D$, the KIE is the ratio $k_H/k_D$.

%%%%%%%%%%%%%
\subsection{\label{sec_Gating}Gating and the classical probability $p(R)$}

Thermal energy from the environment causes the enzyme to vibrate. ``Gating'' \cite{BB92} assumes that this enzyme motion is coupled to the active site configuration by making the donor (C) and acceptor (O) oscillate and sample a range of C-O separations $R_i \leq R \leq R_f$. The likelihood of a separation $R$ occurring is governed by the gating probability distribution $p(R)$. It is a Boltzmann distribution $p(R) = e^{-\beta U^{CO}(R)}/Z^{CO}_{\beta}$ for a potential $U^{CO}(R)$ that describes the sampling of donor-acceptor distances around a mean equilibrium position, $R_e$, at inverse temperature $\beta = 1/(k_B T)$, normalised by a partition function $Z^{CO}_{\beta}$. Assuming a quadratic potential, $U^{CO}(R) = \frac{\mu\Omega^2}{2}(R-R_e)^2$, results in a Gaussian gating probability $p(R)$. Here $\mu$ is the C-O reduced mass and $\Omega$ is the gating frequency. The standard deviation of distances sampled about the peak position $R_e$ is $\sigma = \sqrt{k_B T / (\mu\Omega^2)}$. 

%%%%%%%%%%%%%
\subsection{\label{secParams}Constants and parameters}

Constants for the C-H and O-H Morse potentials are available from standard chemistry data books \cite{Chem_Data}: the experimental dissociation energies 413 kJ/mol and 493 kJ/mol (measured from the zero point energies) are used to derive $D^C$ and $D^O$ (measured from the bottom of the well); the equilibrium distances are $x_e^C=1.09~\AA$ and $x_e^O=0.94~\AA$; and the bond frequencies $\omega^C$ and $\omega^O$ are both $3000 \mathrm{cm}^{-1}$ (in units of wavenumbers). The diatomic well curvatures $a^C$ and $a^O$ are calculated using these values. The squeezing parameter $g$ scales the width of the local electrostatic potentials seen by the transferring particle at the donor (and acceptor) in comparison to the widths of diatomic bonds. If the potentials $V^C$ and $V^O$ in the substrate are narrower in comparison to the isolated C-H and O-H bonds respectively, then the bond frequency increases and mathematically this is reflected in a value of $g>1$. Here we choose $g=2.3$ throughout.
%{\color{red}One expects the Morse potentials to become narrower }and the bond frequency to increase, i.e. $g>1$. 

The quantum transfer rate $\tau_R$ accounts for contributions from transitions between the ground state of the initial potential and various energetic states of the new double well potential. While theoretically all transfers will have a non-zero probability, environmental noise will limit the number of energetic levels that can be reached by the transferring particle. Here we choose to include transfers to the lowest 15 energetic eigenstates of the double well potential. 

The donor-acceptor range $R_i \leq R \leq R_f$ governs which rate contributions are included in the rate Eq.~\eqref{Our_avg_k}. Recall that $R = d + x_e^O + x_e^C$ is determined by the distance $d$ between the $V^C$ and $V^O_R$ well minima. We assume that $d$ cannot be negative, i.e. C and O can be no closer than $R_i = x_e^C+x_e^O = 2.03~\AA$. We choose the maximal value of $d$ to be $3~\AA$ and so $R_f = 5.03~\AA$. While this is the integration range we allow, in the end the probability $p(R)$ ``controls'' the window of C-O separations that are most relevant in the overall transfer rate $k$, see Eq.~\eqref{Our_avg_k}.

\begin{figure*}[t]
\hspace{0.5cm}
\includegraphics[trim=2cm 1cm 3cm 1cm, clip, width=0.98\textwidth]{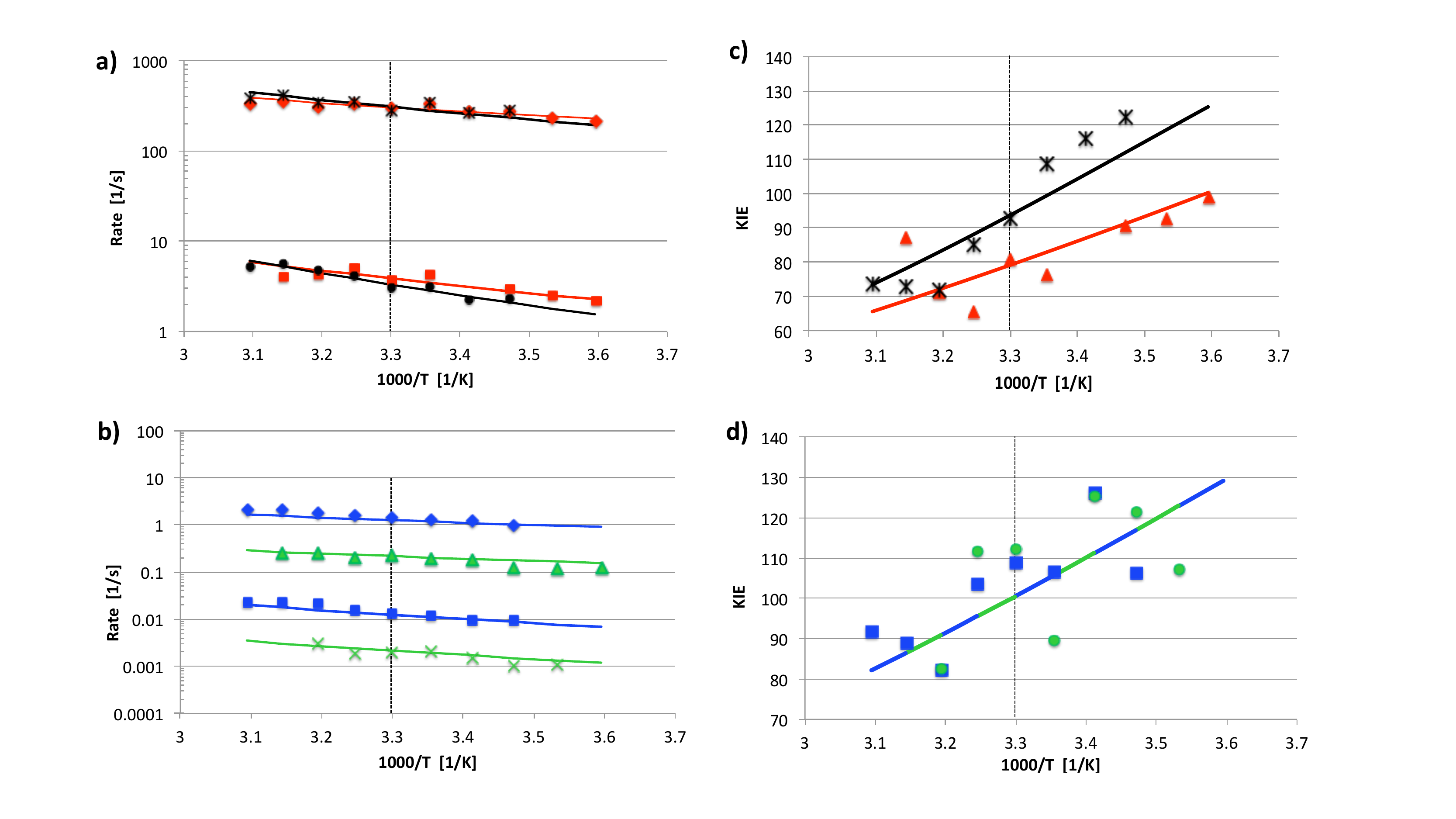}
\caption{\label{fig:ks+KIEs} 
Figures {\bf a)} and {\bf b)} show Arrhenius curves of the transfer rates calculated with Eq.~\eqref{Our_avg_k} for SLO wild-type (WT) and its mutants M1, M2 and M3 (solid lines) using the parameter values for $R_e$ and $\Omega$ given in Table~\ref{table1}. Also displayed are the experimental data  (symbols), cf.  Fig.~\ref{figKRK02}, with the rate values reported in the supplement of [\onlinecite{KRK02}].
{\bf a)} Lines correspond to calculated rates for WT (red) and M3 (black). The upper rates are for hydrogen transfer in WT (red diamonds) and M3 (black crosses), the lower rates are for deuterium transfer in WT (red  squares) and M3 (black circles). 
{\bf b)}  Lines correspond to calculated rates for M1 (blue) and M2 (green). From the top: the rates are for hydrogen transfer in M1 (blue diamonds), hydrogen transfer in M2 (green triangles), deuterium transfer in M1 (blue squares), deuterium transfer in M2 (green crosses). 
Figure {\bf c)} shows the corresponding KIE curves for SLO WT (red) and  M3 (black).
Figure {\bf d)} shows the corresponding KIE curve for M1 and M2 (blue and green dashed). 
The KIE data points obtained from experiment~\cite{KRK02} are also indicated: SLO WT (red triangles) and M3 (black crosses), M1 (blue squares) and M2 (green circles). On all the plots, the temperature at $30^\circ$C is indicated by a dashed vertical line. } 
\end{figure*}

Physically it is reasonable for the isotope-independent rate prefactor $N$ to depend on the temperature as it reflects the environment's impact on the rate dynamics. To obtain a specific functional form will require a more detailed model of the environment, following for example Refs [\onlinecite{BGM10}] and [\onlinecite{SHS_07}]. Here we choose the rate prefactor $N$ to be independent of the isotope (H or D) for each SLO mutant, see Table~\ref{table1}, with its value set by fitting the calculated hydrogen rates to the experimental data. This choice fixes the scale for the hydrogen and deuterium rate plots which are shown in Fig.~\ref{fig:ks+KIEs}{\bf a)} and {\bf b)}. While these plots could be modified by a temperature-dependent $N$, the relative behaviour of rates captured by KIEs, plotted in Fig.~\ref{fig:ks+KIEs}{\bf c)} and {\bf d)}, are unaffected by $N$ even if it is temperature-dependent.

In addition to varying temperature $T$ and isotope mass $m$, we will investigate how mutations of SLO from its wild-type affect the rates and the KIEs predicted with Eq.~\eqref{Our_avg_k}. Since mutants catalyse the same reaction, the functional form of the Morse potentials $V^C, V^O_R$ and $V_R^{CO}$ experienced by the transferring particle remain the same. However, the enzyme mutations will affect the equilibrium donor-acceptor distance, $R_e$, and the gating frequency, $\Omega$, of the gating distribution $p(R)$, and this determines the likelihood that a potential $V_R^{CO}$ will be seen by the transferring particle. The parameter values of $R_e$ and $\Omega$ for the various SLO mutants are discussed in the next section.

%%%%%%%%%%

%%%%%%%%%%

%%%%%%%%%%%%%
\section{\label{sec4} Results}

The hydrogen and deuterium transfer rates, $k_H$ and $k_D$, for SLO WT and each SLO mutant are calculated with Eq.~\eqref{Our_avg_k} using the mutant-specific values for the mean C-O separation $R_e$ and the gating frequency $\Omega$, as listed in Table~\ref{table1}. We first discuss the top four SLO variants, WT, M1, M2 and M3. The values of $R_e$ and $\Omega$ for these are chosen to provide the best fit to the experimental KIE data \cite{KRK02} with gating frequencies in the physically reasonable range $\Omega \leq 400~\mathrm{cm}^{-1}$. 
Figs.~\ref{fig:ks+KIEs} {\bf a)} and {\bf b)} show the calculated H and D rates for the four SLO variants WT and M3 ({\bf a}), and M1 and M2 ({\bf b}), over the experimental temperature range $5^\circ\mbox{C} \leq T \leq  50^\circ\mbox{C}$ together with the observed rates\cite{KRK02}, c.f.  Fig.~\ref{figKRK02}. The corresponding KIEs are shown in Figs.~\ref{fig:ks+KIEs} {\bf c)} and {\bf d)} together with the ratio of the experimentally measured rates as a function of inverse temperature. 

Comparing the experimental and theoretical KIEs we find excellent agreement for the KIE gradient for SLO WT and its mutants M1 and M2. The agreement for the distant mutant M3 is less good than what has been achieved with other models\cite{ESS_H_2010,SH_S_2016,Pollak}. Our calculated M3 deuterium rate does not vary as strongly with temperature as the experimentally observed one. Nevertheless our parameters produce a M3 KIE that is the most temperature dependent of the four SLO variants, WT, M1, M2 and M3, in agreement with experiment.

The SLO WT parameters are $R_e=2.9~\AA $ and $\Omega=400~\mathrm{cm}^{-1}$. Any smaller value of $R_e$ would require an even higher $\Omega$, i.e. result in unrealistically high WT donor-acceptor vibrations. For M3 our equilibrium separation is $R_e = 3.05~\AA$, i.e. $0.15~\AA$ higher than that of WT. The M3 gating frequency is reduced to $\Omega = 325~\mathrm{cm}^{-1}$ and implies an increase in the standard deviation $\sigma$ of the gating in comparison to WT. The SLO WT and M3 gating distributions $p(R)$ at 30$^\circ$C are thus determined by peak and standard deviations $R_e \pm \sigma = 2.9 \pm 0.08~\AA$ and $R_e \pm \sigma = 3.05 \pm 0.10~\AA$, respectively. The values of $R_e$ and $\Omega$ for WT and M3 end up close to the ones reported in Ref. [\onlinecite{SH_S_2016}] where a proton-coupled electron transfer model  was used to derive the rates. There a choice of $R_e = 2.88~\AA$ and $\Omega = 368.2~\mathrm{cm}^{-1}$ for the WT, and $R_e = 3.08~\AA$ and $\Omega = 295.1~\mathrm{cm}^{-1}$ for M3 (see Table 1 of [\onlinecite{SH_S_2016}]), provided a very good fit to the experimental data when an effective mass of $M=10$ amu was chosen for the proton donor-acceptor vibrational mode\cite{SH_S_2016}. Despite the different approaches, the similarity of the parameter values presented here and in [\onlinecite{SH_S_2016}] is particularly noteworthy.

The rates calculated with Eq.~\eqref{Our_avg_k} for SLO mutants M1 and M2 are displayed in Fig.~\ref{fig:ks+KIEs}{\bf b)} and show good agreement with the experimental data, which are also shown. To match the experimental KIEs required an increase of $R_e$ by $0.05~\AA$ in comparison to WT and reduction of the gating frequency $\Omega$  by ca.~5\%. The parameter values for both M1 and M2 that give the best KIE fit are then $R_e = 2.95~\AA$ and $\Omega = 380~\mathrm{cm}^{-1}$. Thus the sampling range increases only very slightly and peak and standard deviation of $p(R)$ at 30$^\circ$C are $R_e \pm \sigma = 2.95 \pm 0.08~\AA$. The calculated KIEs are displayed in Fig.~\ref{fig:ks+KIEs}{\bf d)} together with the experimental data points.

%%%%%%%%%
\begin{table}
\centering
 \begin{tabular}{ | m{0.9cm} ||  m{0.9cm} | m{1.3cm} | m{1.9cm}  | m{0.8cm}  |m{1.3cm} | } 
\hline
    SLO variant &  $R_e$ (\AA) &  $\Omega$ (cm$^{-1}$) & $N$ & KIE & Exp. KIE \\ \hline
    WT	&		2.9	&	400	&	 $4\times 10^{-6}$  & 79 & 81	\\ 
    M1	&		2.95	&	380 	& 	$6\times 10^{-8}$ & 101 & 109\\ 
    M2 	&		2.95	&	 380 	&  	$1.04\times 10^{-8}$ & 101 & 112	\\ 
    M3  	&		3.05 	&	 325	&  	$1\times 10^{-4}$  & 94 & 93	\\ \hline
    DM 	& 	 	3.3 	& 	495.7  & 	Fig.~\ref{fig_KIEs_DM} - grey & 563 & 729  \\ 
    DM 	& 	 	3.8 	& 	185.3  & 	Fig.~\ref{fig_KIEs_DM} - orange & 696 & 729 \\ \hline 
\end{tabular}
\caption{   \label{table1} 
Top: Parameter values for SLO WT and SLO mutants M1, M2 and M3 resulting in reaction rates $k_H$ and $k_D$ and KIEs shown in Fig.~\ref{fig:ks+KIEs}. For these SLO variants the parameters $R_e$ and $\Omega$ are chosen such that the KIEs calculated with Eq.~\eqref{Our_avg_k} give the best fit to the experimental KIE data points \cite{KRK02,SLO_DM}, which are also displayed in Fig.~\ref{fig:ks+KIEs}{\bf c)} and {\bf d)}. The prefactors $N$ have been set so that the calculated hydrogen rates, Fig.~\ref{fig:ks+KIEs}{\bf a)} and {\bf b)}, are as close as possible to experimental data. They set the scale for the rates but cancel out for KIEs.  
The last two columns give the KIE values at $30^{\circ}$C resulting from the model according to Eq.~\eqref{Our_avg_k}, and those calculated from experimental rate data \cite{KRK02,SLO_DM} (last column).
Bottom: For the double mutant (DM) the parameter sets of the flattest (grey) and steepest (orange) KIE curves shown in Fig.~\ref{fig_KIEs_DM} are given, together with the calculated and experimental\cite{SLO_DM} KIE at $30^{\circ}$C.}
\end{table}
%%%%%%%%%%%

%%%%%%%%%%
\begin{figure}[t!]
\includegraphics[trim=0.7cm 0.2cm 0.2cm 0.1cm,clip=true,width=0.45\textwidth]{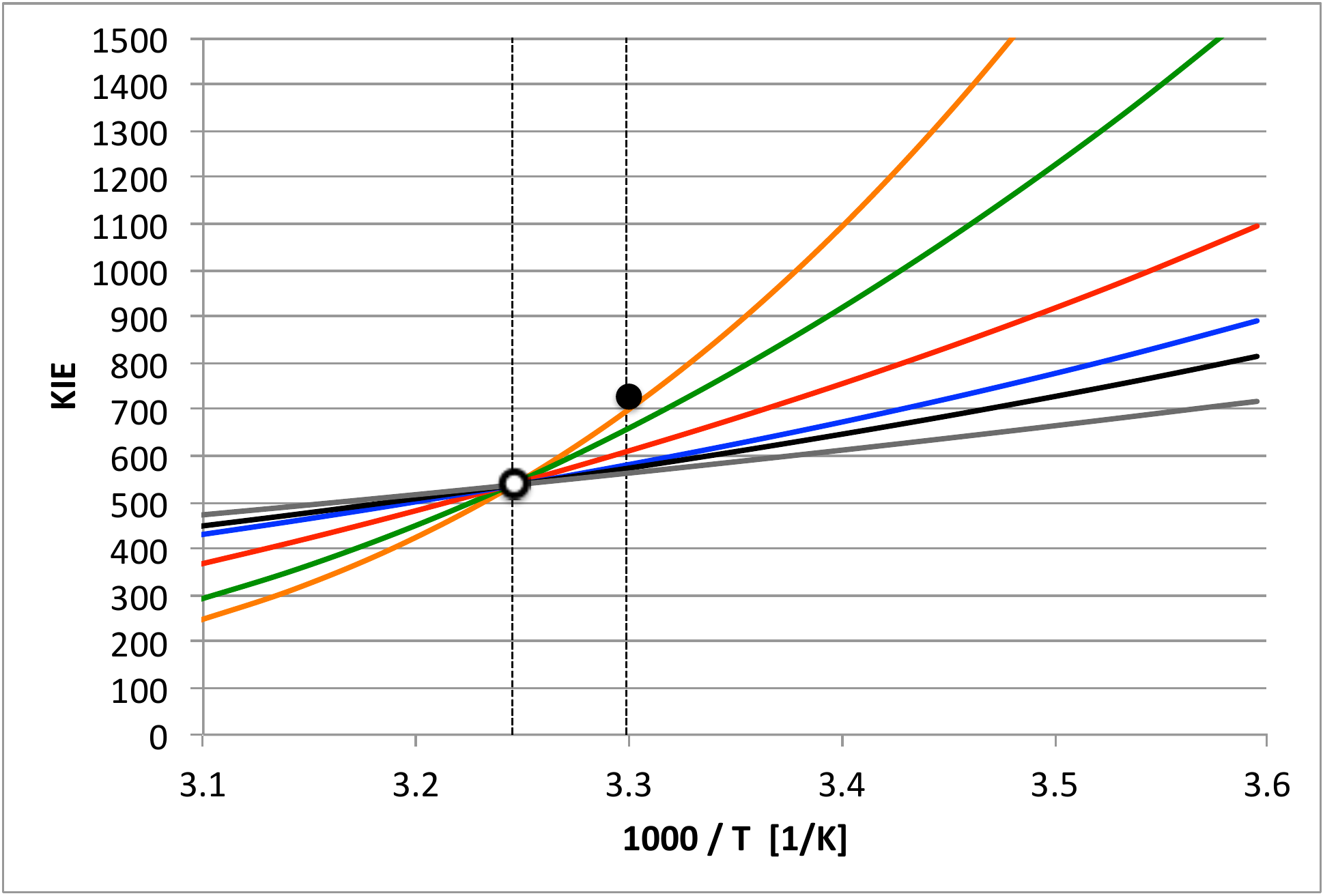}
\caption{KIE plots vs. inverse temperature calculated with Eq.~\eqref{Our_avg_k} for various parameter choices, $R_e$ and $\Omega$, for SLO double mutant (DM) \cite{SLO_DM}. 
All choices are such that they reproduce the measured KIE of $537$ at $35^\circ$C from Ref. [\onlinecite{SLO_DM}].  In order of decreasing $R_e$: orange  $(R_e,\Omega) = (3.8 \, \AA,185.3 \, \mathrm{cm}^{-1})$, green $(R_e,\Omega) = (3.7 \, \AA,199.8 \, \mathrm{cm}^{-1})$, red  $(R_e,\Omega) = (3.6 \, \AA,220.4 \, \mathrm{cm}^{-1})$, blue $(R_e,\Omega) = (3.5\, \AA,256.6 \, \mathrm{cm}^{-1})$, black $(R_e,\Omega) = (3.4\, \AA,326.4\, \mathrm{cm}^{-1})$ and grey $(R_e,\Omega) = (3.3\, \AA,495.7\, \mathrm{cm}^{-1})$.  Dashed lines indicate the temperatures at which the KIEs have been measured\cite{SLO_DM}, $35^\circ$C and $30^\circ$C, and the independent experimental points are shown as filled/unfilled black circles: the unfilled circle is for a KIE of $537\pm 55$ at $35^\circ$C using single-turnover kinetics, and the filled circle is for a KIE of $729\pm 26$ at $30^\circ$C using steady-state measurements. The KIE increase from $35^\circ$C to $30^\circ$C for these curves ranges from ca. $160$ (orange) to ca. $25$ (grey).}
\label{fig_KIEs_DM}
\end{figure}
%%%%%%%%%%

Finally, we compare KIE predictions of the developed model against the KIEs reported in a recent SLO double mutant (DM) experiment\cite{SLO_DM}. Here both the M1 and M2 mutations (Leu$^{546} \rightarrow$ Ala and Leu$^{754} \rightarrow$ Ala) were implemented on a single SLO enzyme. Huge KIEs were observed using two independent methods for the same hydrogen transfer reaction: $537\pm 55$ at $35^\circ$C using single-turnover kinetics and $729\pm 26$ at $30^\circ$C using steady-state measurements. 
Since these are completely different measurements resulting in systematic errors, it is not possible to use these two data points to conclusively infer the temperature dependence of the DM KIE. We thus calculate a set of possible KIE curves with Eq.~\eqref{Our_avg_k}, where each curve has a different parameter pair $(R_e, \Omega)$, see Fig.~\ref{fig_KIEs_DM}. All the curves are pinned at  $537$ at $35^\circ$C, which is the KIE from Ref. [\onlinecite{SLO_DM_2017}] measured using the more reliable steady-state method. 

We find that obtaining a high KIE of 537 at $35^\circ$C requires quite large equilibrium separations, $R_e = 3.3~\AA$ or more, when the gating frequency $\Omega$ is assumed not to exceed $500 \mathrm{cm}^{-1}$. Fig.~\ref{fig_KIEs_DM} shows the calculated KIE curves in the temperature range $5^{\circ}C \leq T \leq 50^{\circ}C$ for parameters in the ranges $3.3~\AA \leq R_e \leq 3.8~\AA$ and $180~\mathrm{cm}^{-1} \leq \Omega \leq 500~\mathrm{cm}^{-1}$. The KIE gradients vary strongly as the parameters are changed. 
At a high value of $\Omega \approx 500~\mathrm{cm}^{-1}$, which corresponds to a very rigid active site, the donor-acceptor separation is $3.3~\AA$ and these values result in a fairly small variation of the KIE with temperature. Allowing $R_e$ to increase to $3.8~\AA$ implies a frequency of $\Omega \approx 190~\mathrm{cm}^{-1}$ and results in a steep KIE increase of ca. 160 between $35^\circ$C and $30^\circ$C.
Calculations for the DM KIE with a different model \cite{SH_S_2016}, have previously suggested an equilibrium separation of $R_e\approx 3.3~\AA$ while the gating frequency was given as $\Omega \approx 280~\mathrm{cm}^{-1}$. This is significantly smaller than the $\Omega \approx 500~\mathrm{cm}^{-1}$ obtained here at the same equilibrium distance, see the grey curve in Fig.~\ref{fig_KIEs_DM}. 
While this manuscriprt was under review, new experimental data were published in Ref.~[\onlinecite{SLO_DM_2017}] that provide further evidence supporting the hypothesis of a SLO DM KIE with a very small temperature dependence.

%%%%%%%%%%%%%
\section{\label{sec5}Tunnelling contribution}

The presented model allows us to determine whether hydrogen tunnelling contributes to the observed rates. Hydrogen starts with a very high probability of over 99\% in the energetic ground state $E_0^C$ of the donor potential $V^{C}$. It is then exposed to the combined donor-acceptor potential $V_{R}^{CO}$. At a given $R$, hydrogen tunnels if its initial energy $E_0^C$ is less than the height of the barrier of the potential $V_{R}^{CO}$. To quantify whether hydrogen tunnelling contributes significantly to the rate $k$ in Eq.~\eqref{Our_avg_k} we identify the distance $\bar{R}$ that contributes most to it, i.e. $\bar{R}$ is the value of $R$ at which the product $p(R) \tau_R$ is maximised. For example, for WT SLO at 30$^\circ$C the distance contributing most to the rate is found to be $\bar{R} = 2.67~\AA$. At this distance the energy difference between barrier height and initial energy is $E_{\rm diff} = V_{\bar{R}}^{CO} (x_b) - E_0^C \sim 0.47 $~eV~$\approx 11$~kcal/mol. We note that this energy barrier is an order of magnitude larger than activation energies obtained from an Arrhenius plot of the experimental data \cite{KRK02}. The chance that hydrogen is thermally excited from the ground state to the top of the barrier is exponentially suppressed and hence it is highly unlikely that hydrogen hops over the barrier. Thus the qualitative model developed here suggests that the dominant transfer mechanism is tunnelling.

%%%%%%%%%%%%%
\section{\label{sec7} Discussion and further work}

% recap
In this paper we investigated the hydrogen transfer reaction catalysed by soybean lipoxygenase. We developed a qualitative model that gives the temperature dependence of the primary KIE through the rate $k$ given in equation \eqref{Our_avg_k}. The model treats the dynamics of the transferring particle quantum mechanically resulting in $\tau_R$, a quantum contribution to the rate $k$. In addition it accounts for the enzyme's role in the transfer via a classical ``gating'' rate, $p(R)$, which arises because of the coupling of the enzyme's vibrations to the donor-acceptor separation. 

% tau_R
% choice of g
The quantum rate $\tau_R$ is fixed by the type of chemical reaction - here the hydrogen (deuterium) transfer catalysed by SLO - and depends principally on the mass of the transferring isotope which will be crucial for the KIE. $\tau_R$ is determined by the double-well potential $V^{CO}_R$. One of the parameters that fixes its shape is the squeezing parameter $g$. If $V^{CO}_R$ were the sum of isolated diatomic C-H and O-H potentials, then $g$ would be 1. 
A larger $g$ makes the individual wells narrower and in our calculation we find that choosing $g=2.3$ gives rate and KIE predictions that are in good agreement with experimental data for SLO WT and all mutants analysed here. This suggests that, when hydrogen (or deuterium) is bonded to the donor or acceptor in the presence of the substrate and the enzyme, then hydrogen experiences a stronger attraction to either C and O  during the rate-limiting step. This slightly higher $g$ value of $2.3$ produces a higher barrier in $V^{CO}_R$ and leads to enlarged KIEs. A similar finding was reported in the intensive computational study of SLO, where the barrier had to be increased manually to obtain KIEs that were as high as the experimental ones \cite{TVLLY06}.

% p(R)
% choice of R and Omega %DM 
With the quantum part of the rate fixed, one is left with the classical gating rate $p(R)$ which is determined by just two free parameters. The donor-acceptor equilibrium separation $R_e$ and the gating frequency $\Omega$ set the average position and spread of the classical gating distribution $p(R)$. It is through changing these two parameters only that we obtain the various rates and KIEs of all the SLO mutants. These two parameters are also the ones that parameterise the KIE curve of each mutant in the non-adiabatic proton-coupled electron transfer reaction model of Ref [\onlinecite{SH_S_2016}] with which we compared our results. A conceptually different semiclassical rate model, based on a Caldeira-Leggett type Hamiltonian that results in a Langevin equation, parameterises the KIE curve of each mutant with only a single parameter, the friction coefficient \cite{Pollak}. The calculated KIE curves are markedly different from the curves reported here, while also showing agreement with the experimental data within the experimental uncertainty.

We found that our model predictions show good agreement with the experimental data for physically reasonable choices of $R_e$ and $\Omega$, see Table~\ref{table1} and Fig.~\ref{fig:ks+KIEs}.
The general picture that emerges from this model is that in WT the active site is compressed and very rigidly held: its low $R_e$ keeps the donor and acceptor very close on average and its high $\Omega$ indicates little movement around the most likely separation $R_e$, with a spread $\sigma$ of less than $0.1~\AA$.  As we saw in Fig.~\ref{fig:ks+KIEs}{\bf c)}  the WT parameter values, which are expected to be close to the ``optimal'' configuration for SLO, lead to KIEs in the range of 65-100 for biological temperatures with moderate temperature dependence. This suggests that $R_e$ and $\Omega$ are very finely tuned in the SLO WT. 

In the mutants, a larger $R_e$ means that the carbon and oxygen are held further apart on average, and this opening of the active site effectively makes the barrier larger. Specifically, while the barrier height most likely seen by the transferring particle in WT is $35.2$~kcal/mol at $R_e = 2.9~\AA$, the most likely barrier height in M1 and M2 is $48.8$~kcal/mol at $R_e = 2.95~\AA$. The higher barrier in M1 and M2 makes the transfer even more difficult and very significantly lowers the rates of these mutants in comparison to WT. As well as an increase in $R_e$, a mutated SLO also has a decreased $\Omega$, implying that the donor and acceptor oscillate from their equilibrium separation over a larger range. The standard deviation $\sigma$ of the sampling distribution $p(R)$ depends on temperature and gating frequency as $\sigma \propto \sqrt{T}/\Omega$. The interplay of larger equilibrium separation and larger gating ranges results in large KIEs that have a much more pronounced temperature dependence than those of WT. 

We note that the quantum rate $\tau_R$ is large for small $R$ and decays rapidly with $R$, thus smaller separations result in much more efficient hydrogen transfer. This means that the separations which contribute significantly to the overall rate $k$ in Eq.~\eqref{Our_avg_k} can be many $\sigma$ smaller than $R_e$. This was observed in section \ref{sec5}, where we found that the separation $\bar{R}$ that contributes most to the rate $k$ can be much shorter than the most likely equilibrium separation $R_e$. This functional dependence makes the KIEs and their temperature dependence  very sensitive to the values of $R_e$ and $\Omega$, particularly as these parameters take on higher values. Generally we find that the KIE increases  when (i) $R_e$ is fixed while $\Omega$ is increased and (ii) when $\Omega$ is fixed but $R_e$ increases. These tendencies found here are in agreement with the conclusions of previous works that have investigated the variation of the KIE with $R_e$ and $\Omega$ in SLO  \cite{ESS_H_2010,SH_S_2016}.

The presented model is qualitative and does not include several physical properties that have been considered elsewhere. The true reaction takes place within a three-dimensional potential landscape\cite{Pollak}, whereas the model presented here considers only a one-dimensional double-well potential in which the hydrogen can move, thus stretching the C-H bond only. Omitting rate contributions from other normal modes, such as bending modes\cite{Pollak}, can lead to an overestimate of the tunnelling contribution. The model presented here is also adiabatic,  including only a single electronic state. This contrasts with other rate models \cite{SH_S_2016} where rate contributions arise from multiple non-adiabatic transfers. Future extensions of the model could address multidimensional potentials and non-adiabatic transfers.

We found that while the model presented here is quite simple in its structure and dependence on the particle mass and temperature, it qualitatively produces the rate/KIE behaviour observed in the  experiments we compared with for various SLO mutants. The model predictions for the temperature dependence of the KIE for the DM open the possibility to further test the validity of the approach and model. 

The values of the physical constants used here, such as the binding energies of hydrogen to donor and acceptor, are specific to SLO and taken from the literature. But it would be straightforward to replace them with relevant constants for other enzyme-catalysed reactions. Future research could thus address the modelling of rates and KIEs of other enzymatic systems that exhibit significant signatures of tunnelling, for instance the Old Yellow Enzyme family of flavoproteins \cite{HPS09} where particular attention is paid to the role of promoting vibrations. The observed KIEs of these enzymes are not as high as SLO, and tend to be more temperature-dependent than SLO. This could be accounted for by our model by, for instance, choosing a higher $R_e$ and lower $\Omega$ than the SLO WT. Applying the quantum model to these enzymes thus provides a fruitful avenue for testing the importance of the gating frequency and transfer distance in enzymatic systems whose KIEs suggest a large tunnelling contribution.

\acknowledgements

It is a pleasure to thank Judith Klinman and her group for their hospitality and many insightful discussions. We are also very grateful to Nigel Scrutton and Sam Hay for their instructive comments, and to Tobias Osborne and his group for their hospitality. SJ is funded by an Imperial College London Junior Research Fellowship. SJ also acknowledges the following grants: ERC grants QFTCMPS; SIQS by the cluster of excellence EXC 201 Quantum Engineering and Space-Time Research; and EPSRC grant EP/K022512/1. JA acknowledges support by the Royal Society and EPSRC (EP/M009165/1). 

\bibliography{references}

\end{document}